\documentclass[aps,prx,twocolumn,showpacs,psfig,superscriptaddress,longbibliography]{revtex4-1}

\usepackage{times}
\usepackage{graphicx}
\usepackage{float}
\usepackage{latexsym,amsmath,amssymb,bm,euscript}
\usepackage{color}
\usepackage{subfigure}
\usepackage{epstopdf}
\usepackage[colorlinks=true,linkcolor=blue,citecolor=blue]{hyperref}
\usepackage{soul}
\usepackage[normalem]{ulem}
\usepackage{mathrsfs}
\usepackage{amsmath}
\usepackage{lettrine}
\usepackage{bbding}
\usepackage{xspace}
\usepackage{textcomp}
\usepackage{textcase}
\usepackage{setspace}
\usepackage{multirow}
\usepackage[version=4]{mhchem}
\usepackage{gensymb}

\def\para{\ensuremath{/\kern -0.8em /}\xspace}

\def\beqn{\begin{eqnarray}}
\def\eeqn{\end{eqnarray}}
\def\beq{\begin{equation}}
\def\eeq{\end{equation}}

\newcommand{\Beq}{\begin{eqnarray*} }
\newcommand{\Eeq}{\end{eqnarray*} }
\newcommand{\Bmat}{\left(\begin{matrix}}
\newcommand{\Emat}{\end{matrix}\right)}

\begin{document}

\title{Dominant Kitaev interaction and field-induced quantum phase transitions in triangular-lattice \ce{KCeSe2}}

\author{Mingtai Xie}
\affiliation{School of Physical Science and Technology, Lanzhou University, Lanzhou 730000, China}
\affiliation{Beijing National Laboratory for Condensed Matter Physics, Institute of Physics, Chinese Academy of Sciences, Beijing 100190, China}

\author{Zheng Zhang}
\email{zhangzheng@iphy.ac.cn}
\affiliation{Beijing National Laboratory for Condensed Matter Physics, Institute of Physics, Chinese Academy of Sciences, Beijing 100190, China}

\author{Weizhen Zhuo}
\affiliation{School of Physical Science and Technology, Lanzhou University, Lanzhou 730000, China}
\affiliation{Beijing National Laboratory for Condensed Matter Physics, Institute of Physics, Chinese Academy of Sciences, Beijing 100190, China}

\author{Wei Xu}
\affiliation{Department of Physics and Astronomy, Shanghai Jiao Tong University, Shanghai 200240, China}

\author{Jinfeng Zhu}
\affiliation{Department of Physics and Astronomy, Shanghai Jiao Tong University, Shanghai 200240, China}

\author{Jan Embs}
\affiliation{Laboratory for Neutron Scattering and Imaging, Paul Scherrer Institute, Villigen, Switzerland}

\author{Lei Wang}
\affiliation{Wuhan National High Magnetic Field Center and School of Physics, Huazhong University of Science and Technology, Wuhan 430074, China}

\author{Zikang Li}
\affiliation{Wuhan National High Magnetic Field Center and School of Physics, Huazhong University of Science and Technology, Wuhan 430074, China}

\author{Huanpeng Bu}
\affiliation{Department of Physics, Southern University of Science and Technology, Shenzhen 518055, China}

\author{Anmin Zhang}
\affiliation{School of Physical Science and Technology, Lanzhou University, Lanzhou 730000, China}

\author{Feng Jin}
\affiliation{Beijing National Laboratory for Condensed Matter Physics, Institute of Physics, Chinese Academy of Sciences, Beijing 100190, China}

\author{Jianting Ji}
\affiliation{Beijing National Laboratory for Condensed Matter Physics, Institute of Physics, Chinese Academy of Sciences, Beijing 100190, China}

\author{Zhongwen Ouyang}
\affiliation{Wuhan National High Magnetic Field Center and School of Physics, Huazhong University of Science and Technology, Wuhan 430074, China}

\author{Liusuo Wu}
\affiliation{Department of Physics, Southern University of Science and Technology, Shenzhen 518055, China}

\author{Jie Ma}
\affiliation{Department of Physics and Astronomy, Shanghai Jiao Tong University, Shanghai 200240, China}

\author{Qingming Zhang}
\affiliation{School of Physical Science and Technology, Lanzhou University, Lanzhou 730000, China}
\affiliation{Beijing National Laboratory for Condensed Matter Physics, Institute of Physics, Chinese Academy of Sciences, Beijing 100190, China}

\begin{abstract}
	Realizing Kitaev interactions on triangular lattices offers a compelling platform for exploring quantum-spin-liquid physics beyond the conventional honeycomb lattice framework.
	Here, we investigate the triangular-lattice antiferromagnet \ce{KCeSe2}, where multiple probes reveal strong magnetic anisotropy suggesting significant Kitaev physics. 
	Through detailed and combined analysis of magnetization, neutron scattering, and thermodynamic experiments, we identify dominant ferromagnetic Kitaev ($K = -1.82$ K) and antiferromagnetic Heisenberg ($J = 1.34$ K) interactions that stabilize a stripe-$yz$ ordered ground state via an order-by-disorder mechanism.
	Magnetic fields applied along the Kitaev bond direction induce two phase transitions at 1.67 T and 3.8 T, consistent with density matrix renormalization group (DMRG) calculations predictions of a progression from stripe-$yz$ to stripe-canted and spin-polarized phases. 
	Near the 1.67 T quantum critical point, enhanced quantum fluctuations suggest conditions favorable for exotic excitations. These results establish \ce{KCeSe2} as a platform for exploring Kitaev physics on triangular lattices.
\end{abstract}
\date{\today}
\maketitle

\textit{Introduction.}---
Kitaev spin interactions~\cite{KITAEV20062,takagi_concept_2019,annurev}, first introduced in an exactly solvable honeycomb-lattice model, exemplify anisotropic, bond-dependent couplings that stabilize a Kitaev spin liquid (KSL) with fractionalized excitations and topological order~\cite{annurev,PhysRevLett.98.247201,jansa_observation_2018}.
Following the scenario proposed by Jackeli and Khaliullin~\cite{jackeli2009Mott}, several honeycomb-lattice magnets with Kitaev interactions have been identified. 
The 5$d$ iridates $A_{2}$\ce{IrO3} ($A$ $=$ Li, Na, Cu)~\cite{PhysRevLett.114.077202,PhysRevLett.110.097204,PhysRevLett.122.167202} were the first candidates for realizing KSL physics. 
Although their ground states exhibit magnetic order, they remain of significant interest for their Kitaev physics.
The 4$d$ compound $\alpha$-\ce{RuCl3}~\cite{doi:10.1126/science.aah6015,doi:10.1126/science.aay5551,do_majorana_2017} has emerged as another prominent candidate,
exhibiting a magnetic field-induced spin-disordered phase widely interpreted as a KSL state.
Other honeycomb-lattice magnets, including \ce{Na2Co2TeO6}~\cite{lin_field-induced_2021,GaotingLin}, and \ce{YbOCl}~\cite{PhysRevResearch.4.033006,PhysRevResearch.6.043061,PhysRevResearch.6.033274}, have drawn considerable interest due to their exotic spin excitations.

While Kitaev interactions have been extensively studied in honeycomb lattice systems, their exploration on triangular lattices has so far been primarily theoretical. The triangular lattice has long served as a fundamental platform for studying geometric frustration, where Anderson first proposed the quantum spin liquid (QSL) concept within the triangular-lattice Heisenberg model~\cite{ANDERSON1973153,doi:10.1080/14786439808206568}. Recent years have seen growing interest in triangular lattice materials like \ce{YbMgGaO4}~\cite{PhysRevLett.115.167203,li_gapless_2015,PhysRevLett.118.107202} and rare-earth chalcogenides, with \ce{NaYbS2}~\cite{baenitz2018NaYbS2,sarkar2019Quantum,wu2022Magnetic,zhuo2024Magnetism} and \ce{NaYbSe2}~\cite{dai2021Spinon,zhang2022Lowenergya} providing extensive experimental evidence supporting a QSL ground state.

The interplay of geometric frustration and bond-dependent Kitaev interactions on the triangular lattice presents a rich theoretical framework for exploring exotic magnetic states~\cite{PhysRevB.89.014414,PhysRevB.95.024421,PhysRevB.92.184416,Li_2015}. However, experimental realizations of materials exhibiting both triangular-lattice geometric frustration and substantial bond-dependent anisotropy have remained elusive. Initial explorations focused on 5d compounds such as \ce{Ba3IrTi2O9}, where ab initio calculations predicted stripy magnetic order near the Kitaev limit~\cite{PhysRevB.92.165108}. Recent studies have expanded this search to 3d/4d transition metal systems. \ce{CoI2}, a van der Waals triangular-lattice antiferromagnet, exhibits substantial magnon decay and level repulsion attributed to bond-dependent anisotropy, with inelastic neutron scattering revealing quantum critical fluctuations in its spiral-ordered state~\cite{kim2023Bonddependent}. \ce{NaRuO2}, describable by a Kitaev-Heisenberg ($K$-$J$) model, provides an even more compelling case with its quantum disordered ground state displaying persistent spin fluctuations and a continuum of excitations~\cite{razpopov_jeff_2023,bhattacharyya_naruo2_2023,ortiz_quantum_2023}. Despite these advances, single crystals of these materials are often unavailable, particularly for \ce{NaRuO2}, limiting detailed characterization of their magnetic interactions.

Ce-based chalcogenides have recently emerged as ideal systems for exploring triangular-lattice Kitaev physics{\color{blue}~\cite{10.21468/SciPostPhys.9.3.041,kulbakov2021Stripeyz,PhysRevB.106.214431,PhysRevB.110.054445,PhysRevLett.133.096703}}, offering several advantages over \ce{NaRuO2}. These materials are generally available as high-quality single crystals (except for $A$\ce{CeO2}) and feature structural flexibility that enables tuning of Kitaev interactions through chemical substitution. For example, \ce{KCeS2}~\cite{10.21468/SciPostPhys.9.3.041,kulbakov2021Stripeyz,PhysRevB.106.214431} and \ce{CsCeSe2}~\cite{PhysRevB.110.054445,PhysRevLett.133.096703} exhibit stripe-$yz$ magnetic order with low-energy excitations associated with strong Kitaev interactions.

\begin{figure*}[ht]
	\includegraphics[angle=0,width=1\linewidth]{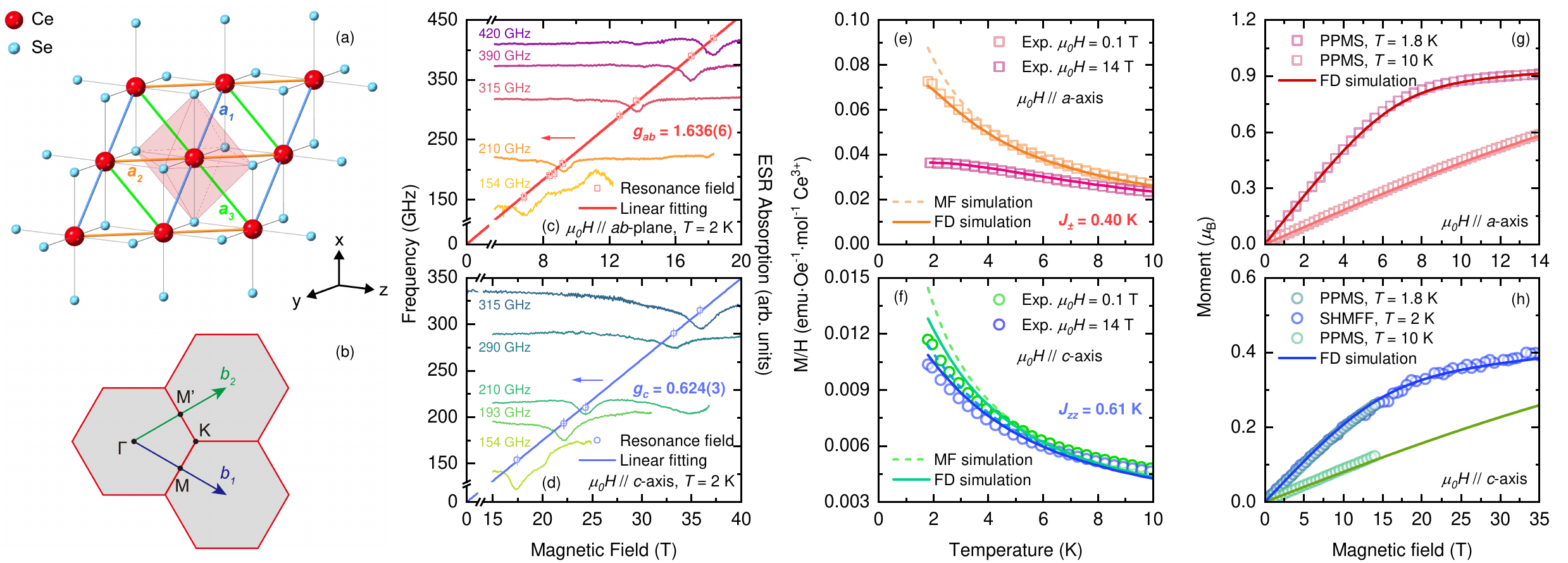}
	\renewcommand{\figurename}{\textbf{Fig. }}
	\caption{ESR and magnetization measurements.
		\textbf{(a)} Triangular lattice of Ce$^{3+}$ ions with Se$^{2-}$ ligands, showing three inequivalent exchange paths ($\vec{a}_{1}$, $\vec{a}_{2}$, $\vec{a}_{3}$).
		\textbf{(b)} Brillouin zone showing high-symmetry points.
		\textbf{(c), (d)} ESR spectra at 2 K showing frequency versus field behavior in $ab$-plane and along $c$-axis. Resonance fields exhibit linear frequency dependence.
		\textbf{(e), (f)} Temperature-dependent magnetization ($M/H$-$T$) under $\mu_0H$ $=$ 0.1~T and 14~T along $a$-axis and $c$-axis, compared with mean-field (MF) and full-diagonalization (FD) calculations.
		\textbf{(g), (h)} Field-dependent magnetization ($M$-$H$) along $a$-axis and $c$-axis at 1.8 K and 10 K (PPMS), and up to 35 T along the $c$-axis at 2 K (SHMFF)~\cite{SM}. Solid lines show FD calculations.
	}\label{FigESR}
\end{figure*}

Here, we present a systematic investigation of \ce{KCeSe2}, synthesized in both single-crystal and polycrystalline forms. Multiple experimental probes reveal strong magnetic anisotropy: electron spin resonance (ESR) measurements show pronounced $g$-factor anisotropy, magnetization data exhibit dominant in-plane response, and neutron diffraction confirms a stripe-$yz$ ordered ground state with highly anisotropic moments. Analysis of inelastic neutron scattering (INS) and magnetization data establishes the presence of competing antiferromagnetic Heisenberg ($J = 1.34$ K) and dominant ferromagnetic Kitaev ($K = -1.82$ K) interactions. Despite the combination of geometric frustration and strong Kitaev interactions typically favoring disordered states, \ce{KCeSe2} exhibits stripe-$yz$ order through an order-by-disorder mechanism. Most notably, applying magnetic fields along the Kitaev bond direction reveals a quantum phase transition at 1.67 T, identified through AC susceptibility measurements. Density matrix renormalization group (DMRG) calculations suggest this transition corresponds to a field-induced quantum critical point characterized by strong quantum fluctuations, before the system ultimately transitions to a fully polarized state at higher fields.

\textit{Model Hamiltonian.}--- 
Ce\textsuperscript{3+} ions with 4$f^{1}$ configuration have a $J$ $=$ 5/2 ground state due to SOC, yielding a theoretical moment of 2.54 $\mu_{\rm B}$ with a Land\'e $g$-factor of $g_{J}$ $=$ 0.9.

In KCeSe\textsubscript{2}, the Ce\textsuperscript{3+} ion is surrounded by six  Se\textsuperscript{2-} anions, forming an octahedral crystal electric field (CEF) environment with $D_{3d}$ point group symmetry.
In this CEF environment, the $J = 5/2$ ground state splits into up to three Kramers doublets, with energy separations typically ranging from tens to hundreds of millielectronvolts~\cite{PhysRevB.103.024430,PhysRevB.104.094421,10.21468/SciPostPhys.9.3.041,PhysRevB.110.054445,PhysRevMaterials.6.084402}.
The low-energy magnetism is governed by the lowest Kramers doublet, reflecting the interplay between SOC and CEF in Ce\textsuperscript{3+} ions.
Raman scattering confirms the first CEF excitation level of \ce{KCeSe2} at $\sim$ 39 meV ($\sim$ 450 K)~\cite{SM}, ensuring that CEF excitations can be neglected when studying the ground state and finite-temperature magnetism below this energy scale.
Thus, effective spin-1/2 operators are constructed from the CEF ground state wavefunctions~\cite{PhysRevLett.115.167203,PhysRevB.94.035107}.
Taking into account the magnetic anisotropy and triangular lattice symmetry, an anisotropic effective spin-1/2 Hamiltonian is developed to describe the ground state and low-energy spin excitations of KCeSe\textsubscript{2}~\cite{PhysRevLett.115.167203,PhysRevB.94.035107},
\begin{align}
	{\mathcal H} = & \sum_{\langle ij\rangle}\left[J_{zz} S_i^z S_j^z+J_{\pm}\left(S_i^{+} S_j^{-}+S_i^{-} S_j^{+}\right)\right. \notag\\
	& +J_{\pm\pm}\left(\gamma_{ij} S_i^{+} S_j^{+}+\gamma_{ij}^* S_i^{-} S_j^{-}\right) \notag\\
	& \left.-\frac{iJ_{z\pm}}{2}\left(\gamma_{ij} S_i^{+} S_j^z-\gamma_{ij}^* S_i^{-} S_j^z+\langle i \leftrightarrow j\rangle\right)\right], \label{eq1:Hamiltonian}
\end{align}
where $S_i^\alpha (\alpha = x,y,z)$ are the effective spin-1/2 operators at site $i$, and $S_i^{\pm} = S_i^x \pm iS_i^y$ are non-Hermitian ladder operators. 
The nearest-neighbor (NN) anisotropic spin-exchange interactions are denoted by $J_{\pm}$, $J_{zz}$, $J_{\pm\pm}$, and $J_{z\pm}$. 
The phase factor $\gamma_{ij}$ takes values of 1, $e^{i \frac{2\pi}{3}}$, and $e^{-i \frac{2\pi}{3}}$ along the three bonds $\vec{a}_1$, $\vec{a}_2$, and $\vec{a}_3$, respectively, as illustrated in Fig.~\ref{FigESR}{\color{blue}(a)}.

The anisotropic Hamiltonian, described in the Cartesian coordinate system,
can be transformed into the crystallographic coordinate frame and reformulated as an extended Kitaev-Heisenberg ($K$-$J$) model~\cite{PhysRevX.9.021017,PhysRevB.98.054408}, thereby facilitating the interpretation of Kitaev physics on triangular lattices.
The parameters conversion between the two modes are given by the following expressions~\cite{PhysRevX.9.021017}:
\begin{equation}
	\begin{array}{l}
		J = \frac{1}{3}\left( 4J_{\pm} + J_{zz} + 2J_{\pm\pm} + \sqrt{2}J_{z\pm} \right)\\
		\\
		K = -2J_{\pm\pm} - \sqrt{2}J_{z\pm}\\
		\\
		\mathit{\Gamma}  =  \frac{1}{3}\left( -2J_{\pm} + J_{zz} - 4J_{\pm\pm} + \sqrt{2}J_{z\pm} \right)\\
		\\
		\mathit{\Gamma'} =  \frac{1}{6}\left( -4J_{\pm} + 2J_{zz} + 4J_{\pm\pm} - \sqrt{2}J_{z\pm} \right)
	\end{array}
	\label{transformation}
\end{equation}
Here, $J$ is the isotropic Heisenberg interaction, $K$ is the anisotropic Kitaev interaction, and $\mathit{\Gamma}$ and $\mathit{\Gamma'}$ are the off-diagonal terms arising from the distorted octahedral structure, as illustrated in Fig.~\ref{FigESR}{\color{blue}(a)}.
In a corner-sharing configuration with a 180$^\circ$ bond angle, $K$ vanishes, whereas in an edge-sharing configuration with a 90$^\circ$ bond angle, $J$ is suppressed~\cite{TREBST20221}. 
In \ce{KCeSe2}, the bond angle is 93.47$^\circ$, close to a right angle, suggesting that Kitaev interactions are dominant. 
This suggests that the \ce{CeSe6} octahedron is nearly regular, effectively reducing the off-diagonal terms $\mathit{\Gamma}$ and $\mathit{\Gamma'}$.
Next, we will discuss the magnetism arising from magnetic anisotropy and Kitaev physics, based on the spin models described above.


\textit{Magnetic anisotropy and stripe-$yz$ order.}---
\ce{KCeSe2} exhibits strong magnetic anisotropy in both its paramagnetic and ground states, as revealed through complementary electron spin resonance (ESR), magnetization, and neutron diffraction measurements.
We employed a high-frequency pulsed-field ESR spectrometer to determine the $g$-factors of the \ce{KCeSe2} single crystal (see Supplementary Materials (SM)~\cite{SM} for details).
Fig.~\ref{FigESR}(c) and (d) display distinct ESR resonance peaks of \ce{KCeSe2} at 2 K for various excitation frequencies $\nu$ in the $ab$-plane and along the $c$-axis, respectively.
Using the resonance condition $\nu = g_{\alpha}\mu_{B}\mu_{0}H_{\alpha}/h$, where $\mu_{B}$, $\mu_{0}$, and $h$ are Bohr's magneton, vacuum permeability, and Planck's constant, respectively, and $\alpha = a, b, c$ represents the crystallographic axis directions of the single-crystal sample, we extracted the $g$-factors: $g_{ab} = 1.636(6)$ for the $ab$-plane and $g_{c} = 0.624(3)$ along the $c$-axis. 
The significant $g$-factor anisotropy ($g_{ab}/g_{c} \simeq 2.6$) indicates strong easy-plane magnetic anisotropy in \ce{KCeSe2}.

Magnetization measurements further confirm the magnetic anisotropy of \ce{KCeSe2}. 
As shown in Fig.~\ref{FigESR}(e) and (f), temperature-dependent magnetization ($M/H$-$T$) demonstrates that magnetization along the $a$-axis is approximately 10 times larger than that along the $c$-axis. 
Mean-field (MF) and full-diagonalization (FD) simulations of $M/H$-$T$ data yield exchange parameters $J_{\pm}$ $=$ 0.40 K and $J_{zz}$ $=$ 0.61 K. The FD calculations, which include $J_{\pm\pm}$ and $J_{z\pm}$ interactions omitted in MF theory, show superior agreement with experimental data at low temperatures and fields. Field-dependent magnetization curves at 1.8 K and 10 K are well reproduced by FD calculations along both crystallographic directions.

\begin{figure}[t!]
	\includegraphics[angle=0,width=1\linewidth]{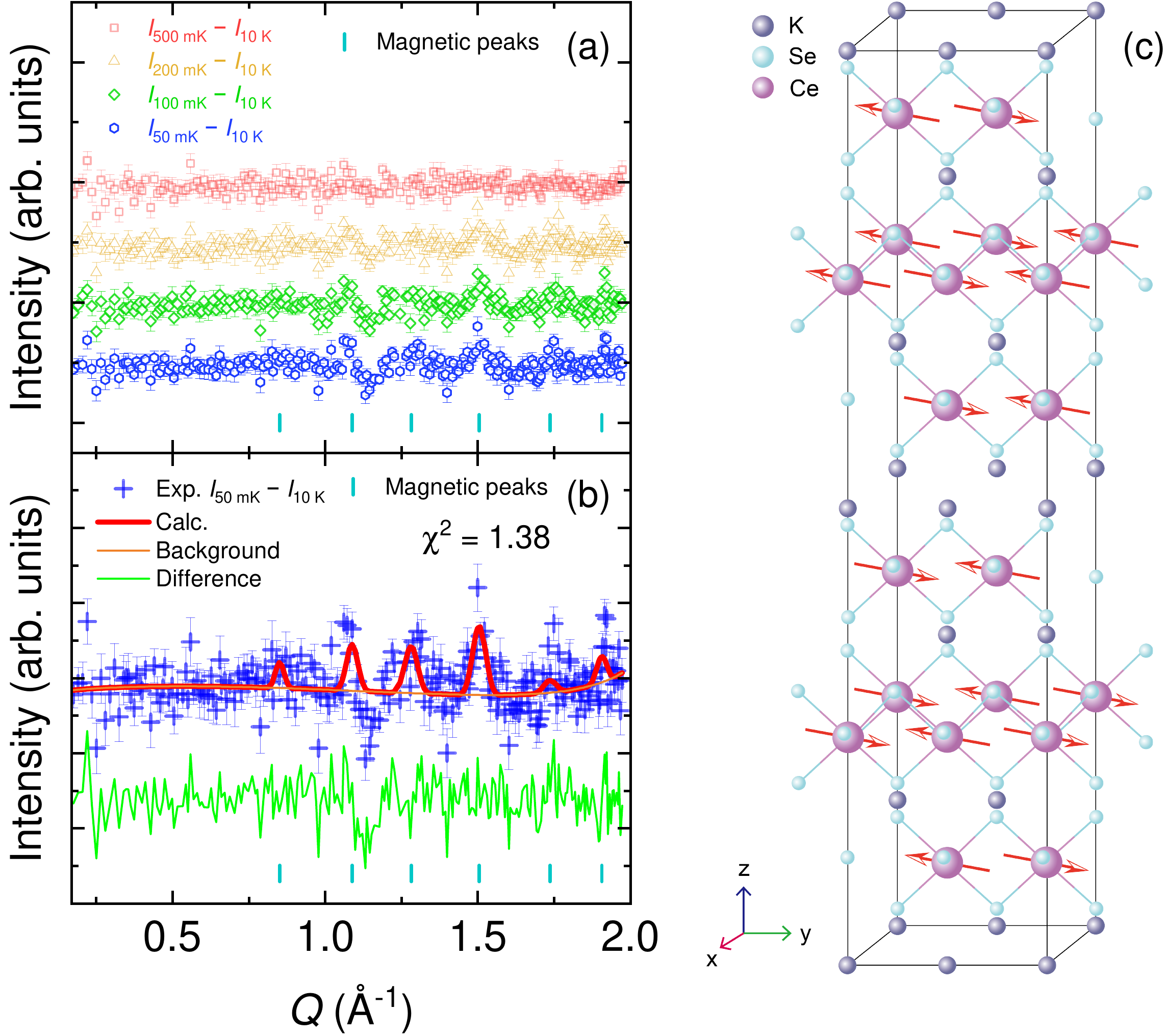}
	\renewcommand{\figurename}{\textbf{Fig. }}
	\caption{Magnetic structure and neutron diffraction analysis.
	\textbf{(a)} Temperature evolution of elastic neutron scattering of \ce{KCeSe2} powder samples. Data shown as difference patterns ($I_{T} - I_{10\rm \ K}$) to highlight magnetic Bragg peaks. Magnetic peak positions are marked by cyan bars.
	\textbf{(b)} Detailed refinement of the 50 mK neutron diffraction pattern (blue crosses) using the stripe-$yz$ magnetic structure model (red line). Background contribution (pink line) and difference plot (green line) are shown. Cyan bars indicate magnetic peak positions.
	\textbf{(c)} Ground state magnetic structure of \ce{KCeSe2} with propagation vector $\vec{Q} = (0, -1/2, 1/2)$. Red arrows indicate ordered Ce$^{3+}$ moments, arranged in a stripe pattern within the $yz$-plane.
	}\label{FigMagneticStructure}
\end{figure}

To investigate the ground-state magnetic anisotropy, we conducted elastic neutron scattering~\cite{neutron} on polycrystalline \ce{KCeSe2} down to 50 mK (\mbox{Fig.~\ref{FigMagneticStructure}}{\color{blue}(a)}). 
Magnetic Bragg peaks were isolated by subtracting the data collected at 10 K, which was used as the background.
Six magnetic Bragg peaks emerge as the temperature decreases.
Refinement of the 50 mK data ($\chi^{2}$ = 1.38, $R_{F,\text{Mag}}$ = 15.76\%) reveals a stripe-$yz$ magnetic structure with propagation vector $\vec{Q} = (0, -1/2, 1/2)$, as illustrated in Fig.~\ref{FigMagneticStructure}{\color{blue}(c)}.
Our refinement indicates that the \ce{Ce^{3+}} moments are predominantly confined to the $ab$-plane ($m_{ab} = 0.68(11) \mu_{\rm B}$), with a negligible out-of-plane component ($m_{c} = 0.050(8) \mu_{\rm B}$).
This magnetic structure breaks the $C_{3}$ rotational symmetry, akin to that observed in \ce{KCeS2}~\cite{kulbakov2021Stripeyz} and \ce{CsCeSe2}~\cite{PhysRevB.110.054445}.
We also attempted to refine the elastic neutron data using the stripe-$x$ configuration as the ground state of \ce{KCeSe2}, which yielded a $R_{F,mag}$=32.69\% (see SM~\cite{SM}), significantly larger than that obtained for the stripe-$yz$ configuration. This further supports the conclusion that the ground state of \ce{KCeSe2} is of the stripe-$yz$ type.

Remarkably, most Ce-based magnets isostructural to \ce{KCeSe2} exhibit a stripe-$yz$ ground state~\cite{kulbakov2021Stripeyz,PhysRevB.106.214431,PhysRevB.110.054445,PhysRevLett.133.096703}. Phase diagrams derived from the Hamiltonian \ref{eq1:Hamiltonian}~\cite{zhu2018Topography,PhysRevX.9.021017} show that stripe-$yz$ order is stabilized in regions where both $J_{\pm\pm}$ and $J_{z\pm}$ are positive. According to the transformation relationship in Eq.~\ref{transformation}, this parameter regime corresponds to ferromagnetic Kitaev interactions. Given the strong magnetic anisotropy observed in our measurements, a quantitative determination of these interaction parameters becomes crucial. This motivates our investigation of the low-energy spin excitations that reflect the underlying exchange couplings.

\begin{figure}[t]
	\includegraphics[angle=0,width=1\linewidth]{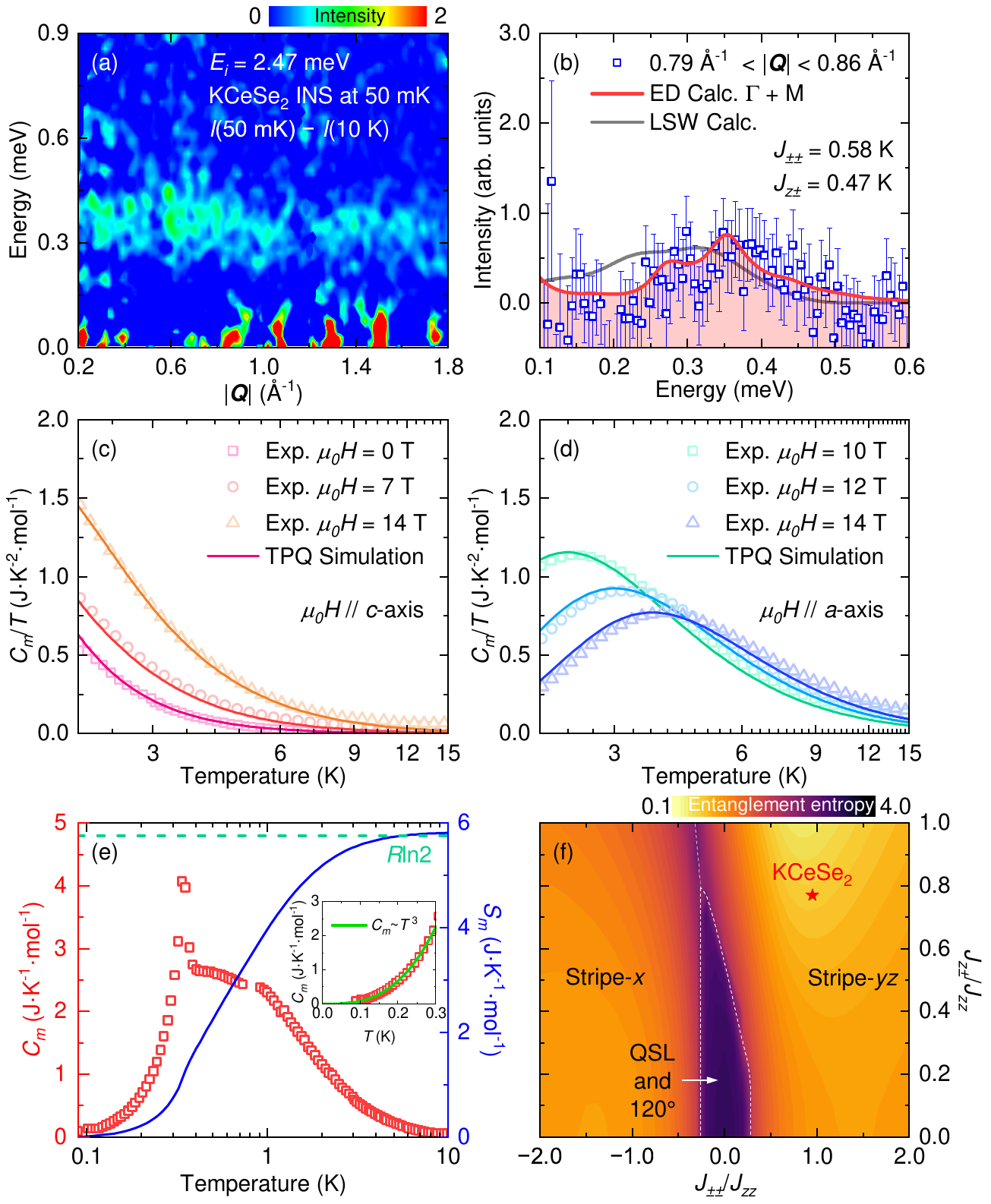}
	\renewcommand{\figurename}{\textbf{Fig. }}
	\caption{Spin dynamics, thermodynamics, and phase diagram of \ce{KCeSe2}.
			\textbf{(a)} INS spectrum of polycrystalline \ce{KCeSe2} measured with incident energy $E_i = 2.47$ meV. Data shown as difference pattern $I$(50 mK) $-$ $I$(10 K) to highlight magnetic excitations.
			\textbf{(b)} Energy dependence of scattered intensity integrated over $|\vec{Q}|$ $=$ 0.79-0.86 \AA$^{-1}$ (blue squares). Red shading shows ED calculations at the $\Gamma$ and M points, while the gray line indicates linear spin wave (LSW) calculations.
			\textbf{(c), (d)} Magnetic specific heat ($C_{m}/T$) under various fields applied along $c$-axis and $a$-axis above 2 K. Experimental data (symbols) compared with thermal pure quantum (TPQ)  state calculations (solid lines).
			\textbf{(e)} Zero-field magnetic specific heat from 0.09-10~K showing a transition at $T_{\text{N}} \sim 0.34$ K. Blue line shows magnetic entropy. Inset: Low-temperature $C_{m}$ follows $T^{3}$ behavior below 0.3~K.
			\textbf{(f)} Ground state phase diagram of Hamiltonian \ref{eq1:Hamiltonian}, obtained from DMRG calculations using experimentally determined diagonal exchange parameters for \ce{KCeSe2} ($J_{\pm}=0.40$ and $J_{zz}=0.61$), with phases distinguished by entanglement entropy. Axes are normalized by $J_{zz}$. White dashed lines indicate phase boundaries. The red star marks the \ce{KCeSe2}'s position.
	}\label{FigSpinExcitation}
\end{figure}

\textit{Low-energy Spin Excitations.}---
To investigate the low-energy spin excitations and Kitaev physics in \ce{KCeSe2}, we performed INS~\cite{neutron} on polycrystalline samples (Fig.~\ref{FigSpinExcitation}(a)) with incident energy $E_i = 2.47$ meV and energy resolution $\delta E_0 \simeq 0.15$ meV (see SM~\cite{SM}).
Subtracting the 10 K background from the 50 mK data reveals a weak but distinct magnetic excitation in the energy range of 0.3 to 0.4 meV.
The spectrum exhibits maximum intensity at $|\vec{Q}| \sim 0$, gradually weakening and becoming more diffuse at higher $|\vec{Q}|$.
The observed dispersion closely resembles that of \ce{KCeS2}~\cite{PhysRevB.106.214431} and \ce{CsCeSe2}~\cite{PhysRevLett.133.096703}, both of which exhibit stripe-$yz$ ground states, but it contrasts sharply with \ce{KCeO2}~\cite{PhysRevB.104.094421}, where excitations peak near the $\rm K$ point.
To further analyze the spin interactions, we examined the INS spectra integrated over $|\vec{Q}|$ $=$ 0.79-0.86 \AA$^{-1}$ (Fig.~\ref{FigSpinExcitation}(b)), encompassing the $(0, 0, 3)$ point ($\Gamma$), $(1/2, 0, 0)$ point ($\rm M$), and $(0, 1/2, 0)$ point ($\rm M'$) in the Brillouin zone.
We employed both exact diagonalization (ED)~\cite{KAWAMURA2017180,ido2024update} and linear spin wave (LSW)~\cite{Toth_2015} calculations with varying exchange parameters to model the spectral features. For the ED calculations, we used a $L_y \times L_x = 4 \times 6$ spin lattice with periodic boundary conditions (PBC) (see SM~\cite{SM}). The calculated spectral weights from ED (red shading in Fig.~\ref{FigSpinExcitation}(b)) with parameters $J_{\pm\pm} = 0.58$~K and $J_{z\pm} = 0.47$~K reproduce the experimentally observed features, including the weak peak at 0.28 meV and the stronger peak at 0.35 meV. The LSW calculations (details in SM~\cite{SM}) capture the overall energy range but show less accuracy in reproducing the detailed spectral weight distribution.
Combined with the previously determined diagonal exchange parameters from magnetization data, all spin-exchange interactions in the Hamiltonian \ref{eq1:Hamiltonian} are now experimentally determined: $J_{\pm} = 0.40$ K, $J_{zz} = 0.61$ K, $J_{\pm\pm} = 0.58$ K, $J_{z\pm} = 0.47$ K.

The determined interactions facilitate a quantitative analysis of the specific heat of \ce{KCeSe2}.
Using the thermal pure quantum (TPQ) state~\cite{KAWAMURA2017180,ido2024update}, we calculate $C_{m}/T$ on a 4$\times$6 spin lattice with PBC under different magnetic fields applied along $c$-axis and $a$-axis.
The excellent agreement between the calculated and experimental data (Fig.~\ref{FigSpinExcitation}(c), (d)) validates the reliability of these interactions. 
Extending this analysis to lower temperatures, specific heat measurements down to 90 mK (Fig.~\ref{FigSpinExcitation}(e)) reveal $\sim T^{3}$ scaling below $T_{\rm N} \sim 0.34$ K (inset), which is characteristic of well-defined magnon excitations in the magnetically ordered state.

Using density matrix renormalization group (DMRG) on a 6$\times$30 spin lattice with cylindrical boundary conditions (CBC)~\cite{ITensor,ITensor-r0.3}, we constructed a ground-state phase diagram (Fig.~\ref{FigSpinExcitation}(f)).
Entanglement entropy identifies distinct phases, placing \ce{KCeSe2} in the stripe-$yz$ region.

The fitted spin-exchange interactions in Hamiltonian \ref{eq1:Hamiltonian} map onto the extended $K$-$J$ model, $J = 1.34$ K, $K = -1.82$ K, $\mathit{\Gamma} = -0.62$ K, and $\mathit{\Gamma'} = 0.21$ K.
The dominant interactions are the ferromagnetic Kitaev interaction $K$ and antiferromagnetic Heisenberg interaction $J$ , along with a noticeable symmetric off-diagonal exchange $\mathit{\Gamma}$.
In the theoretical phase diagram of the $J-K-\mathit{\Gamma}$ model~\cite{PhysRevB.103.054410}, these parameters also place \ce{KCeSe2} clearly within the stripe-$yz$ ordered phase, consistent with our neutron diffraction results.
Remarkably, the emergence of stripe-$yz$ order in \ce{KCeSe2} is counterintuitive. Typically, geometric frustration arising from antiferromagnetic Heisenberg interactions on triangular lattices and exchange frustration induced by Kitaev interactions individually enhance quantum fluctuations, favoring spin-disordered or spin-liquid states. However, in \ce{KCeSe2}, these two sources of frustration combine to surprisingly promote magnetic ordering, suggesting that Kitaev interaction plays a key role in resolving the degeneracies inherent in frustrated triangular lattice magnets. Such ordering via frustration-induced degeneracy lifting aligns with the order-by-disorder scenario~\cite{PhysRevB.98.054411}. This naturally motivates investigating whether perturbing or weakening the Kitaev interaction experimentally could restore a spin-disordered state.

\begin{figure}[t!]
	\includegraphics[angle=0,width=1\linewidth]{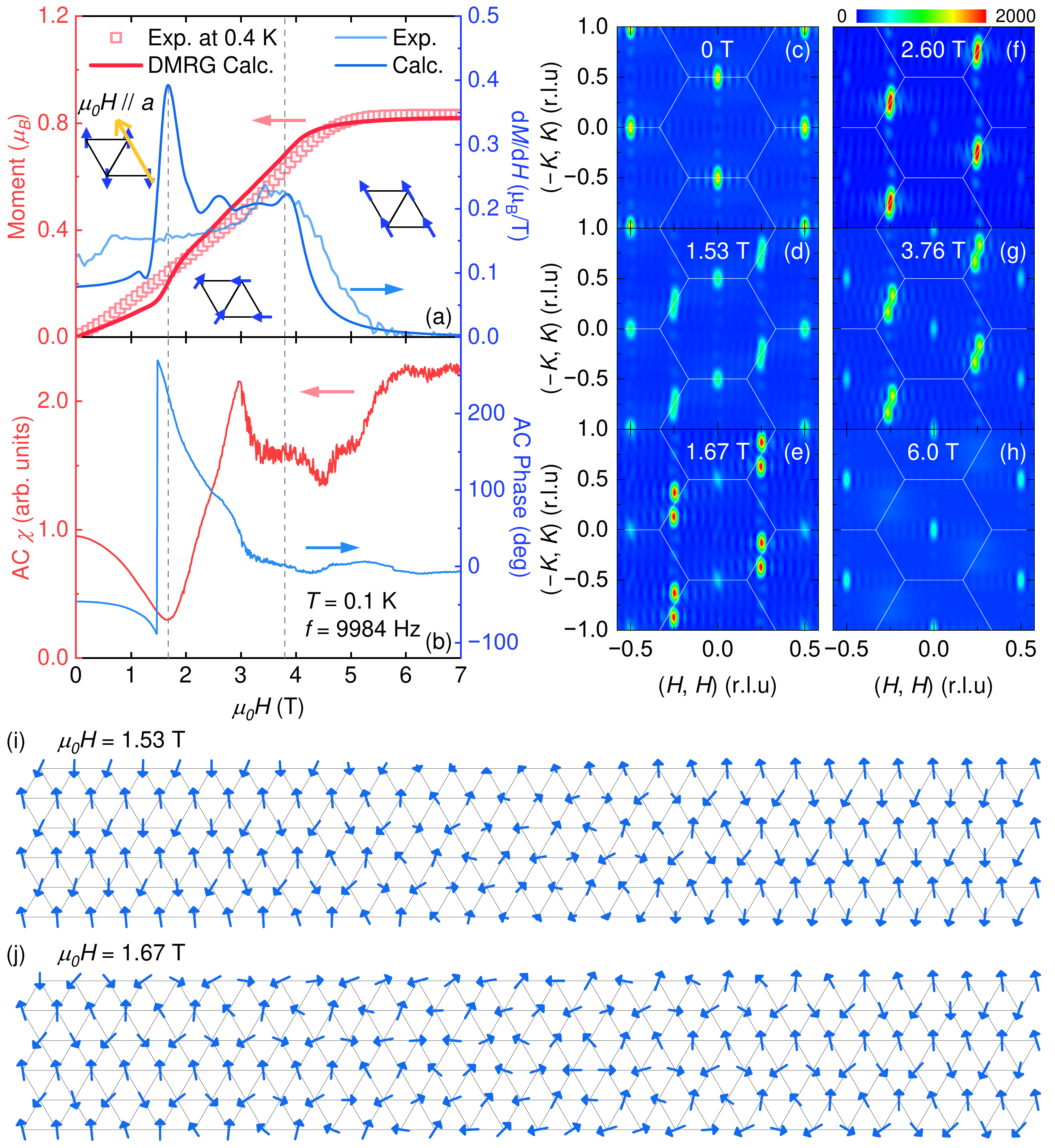}
	\renewcommand{\figurename}{\textbf{Fig. }}
	\caption{Field-induced magnetic phase transition.
		\textbf{(a)} Field-dependent magnetization along $a$-axis at 0.4 K (red squares: experiment; red line: DMRG calculations). $dM/dH$ curves from experiment (light blue) and DMRG (dark blue) show phase transitions. Inset: magnetic field orientation relative to triangular lattice.
		\textbf{(b)} AC susceptibility ($\mathit{\chi}$, red) and phase (blue) measured along $a$-axis at 0.1 K with $f$ $=$ 9984 Hz. Gray dashed lines in (\textbf{(a)}, \textbf{(b)}) mark field-induced phase transitions.
		\textbf{(c)-(h)} Static spin structure factor $S(\vec{Q})$ under different magnetic fields along the \textit{a}-axis.
		\textbf{(i)}, \textbf{(j)} Real-space moment arrangement near QCP from 6$\times$30 DMRG calculations.
	}\label{FigMagneticField}
\end{figure}

\textit{Magnetic field induced phase transition.}---
As a key tuning parameter, magnetic fields play an essential role in advancing the study of Kitaev-related physics in triangular or hexagonal lattice system. For example, in \ce{CsCeSe2}~\cite{PhysRevLett.133.096703}, the application of an external magnetic field can tune the interaction between single-magnon states and the two-magnon continuum, leading to the decay of coherent magnon excitations, level repulsion, and a transfer of spectral weight into the continuum. In the more extensively studied compound $\alpha$-\ce{RuCl3}~\cite{doi:10.1126/science.aay5551,PhysRevLett.119.037201,PhysRevResearch.4.043024}, magnetic fields are known to induce a spin-disordered (or QSL) phase. Motivated by these insights, we explored the effect of an external magnetic field on the stripe-$yz$ ordered state in \ce{KCeSe2}.
Applying a magnetic field along the $a$-axis (parallel to a Kitaev bond direction) provides a controllable means to modify the magnetic ground state via the Zeeman effect.

Initial magnetization measurements at 0.4 K (red squares, Fig.~\ref{FigMagneticField}(a)) reveal two features in $dM/dH$: a broad peak near 3.76 T associated with spin polarization, and a weaker anomaly near 0.7 T suggesting a possible phase transition.
To investigate the field-induced phase transition, we conducted AC susceptibility down to 0.1 K.
Fig.~\ref{FigMagneticField}(b) reveals a sharper transition near 1.67 T, accompanied by a phase flip in the AC response (blue line).
This phase flip provides direct evidence of a transition between distinct magnetic states.

To gain a deeper understanding of the field-induced transitions, we performed DMRG calculations  on a 6$\times$30 lattice with CBC. 
The calculated $M$-$H$ and $dM/dH$ curves (red and blue lines in Fig.~\ref{FigMagneticField}(a)) align closely with experimental results, accurately capturing the polarization field at 3.76 T and the 1.67 T phase transition observed in AC susceptibility.
The matrix product states (MPS) obtained from DMRG calculations enable direct visualization of the magnetic structure directly in both real and momentum space under different fields. 
The static spin structure factor (SSSF) under different fields (Fig.~\ref{FigMagneticField}(c)-(h)) reveals a peak at $\rm M$ point below 1.67 T, indicative of stripe-$yz$ order. 
Near the critical field, quantum fluctuations weaken the $\rm M$-point peak and introduce double peaks near $\rm M'$-point.
Beyond the critical field of 1.67 T, the $\rm M$ point peak disappears, while the peak at $\rm M'$ becomes more pronounced.
As shown in Fig~\ref{FigMagneticField}(f), the arrangement of magnetic moments is ordered, forming a stripe-canted phase.
As the magnetic field continues to increase, the spin system transitions toward a polarized state, as evidenced by the diminishing $\rm M’$ peak and the increasing intensity at the $\Gamma$ point in the SSSF.

The behavior near the QCP of 1.67 T is particularly intriguing. 
On the one hand, the SSSF exhibits multi-$Q$ characteristics (Fig~\ref{FigMagneticField}(d, e)); on the other hand, the arrangement of magnetic moments appears to be disordered (Fig~\ref{FigMagneticField}(i, j)).
Considering the geometric frustration of the triangular lattice and the significant Kitaev interactions in \ce{KCeSe2}, the magnetic-field-driven quantum phase transition introduces strong quantum fluctuations that destabilize the previously well-defined magnons. This may potentially lead to the emergence of deconfined spinon excitations near the quantum critical field. Although further dynamical studies are needed to confirm this scenario, the sharp changes observed in both AC susceptibility and AC phase near the critical field, together with the nearly disordered features revealed by our DMRG calculations, provide preliminary evidence for the rich quantum phenomena associated with the QCP.
We also notice that in \ce{CsCeSe2}, applying a magnetic field along the $a$-axis induces significant changes in low-energy spin excitations at 2.3 T, exhibiting a continuous-like dispersion behavior compared to zero field~\cite{PhysRevLett.133.096703}. 
This suggests that \ce{CsCeSe2} possibly undergoes a field-induced phase transition analogous to that observed in \ce{KCeSe2}.

\textit{Conclusions.}---
We have conducted comprehensive investigations of quantum magnetism in \ce{KCeSe2} through magnetic and thermodynamic measurements combined with numerical calculations. Multiple experimental probes reveal strong magnetic anisotropy: ESR shows pronounced $g$-factor anisotropy, magnetization exhibits dominant in-plane response, and neutron diffraction confirms a stripe-$yz$ ordered ground state with highly anisotropic moments. This anisotropic behavior suggests significant Kitaev interactions, which we quantify through detailed analysis of magnetization, heat capacity, and INS data, revealing dominant ferromagnetic Kitaev ($K = -1.82$~K) and antiferromagnetic Heisenberg ($J = 1.34$~K) interactions.
The interplay of Kitaev interactions and geometric frustration stabilizes the stripe-$yz$ order through an order-by-disorder mechanism. 
Furthermore, applying magnetic fields along the $a$-axis reveals a quantum phase transition near 1.67~T, where DMRG calculations suggest a field-induced strong quantum fluctuation regime. We report the first observation of field-induced phase transitions in a stripe-$yz$ ordered triangular magnet, with the quantum critical region near 1.67~T offering new opportunities to explore exotic magnetic physics under competing interactions. These findings establish \ce{KCeSe2} as an important platform for studying Kitaev physics on triangular lattices.

\textit{Acknowledgements.}---
This work was supported by 
the National Key Research and Development Program of China (Grant Nos. 2024YFA1408300 and 2022YFA1402704), 
the National Science Foundation of China (Grant No. 12274186), 
the Strategic Priority Research Program of the Chinese Academy of Sciences (Grant No. XDB33010100), 
and the Synergetic Extreme Condition User Facility (SECUF, \href{https://cstr.cn/31123.02.SECUF}{https://cstr.cn/31123.02.SECUF} ).
We acknowledges the support from Users with Excellence Program of Hefei Science Center and High Magnetic Field Facility, CAS.

%

\end{document}